\RequirePackage{fix-cm}
\documentclass[smallextended]{svjour3}       
\smartqed  
\usepackage{graphicx}
\usepackage{latexsym}
\usepackage{cite}
\usepackage{array}
\usepackage{caption}
\usepackage[footnotesize]{subfigure}
\usepackage{stfloats}
\journalname{Neural Computing and Applications}

\begin{document}

\title{Multidimensional Urban Segregation
}
\subtitle{Toward A Neural Network Measure}


\author{Madalina Olteanu\and Aur\'elien Hazan\and Marie Cottrell\and Julien Randon-Furling}


\institute{M. Cottrell \at
              SAMM (EA4543), Universit\'e Paris 1 Panth\'eon-Sorbonne, Paris, France\\
              \email{Marie.Cottrell@univ-paris1.fr}           
           \and
           A. Hazan \at
              LISSI (EA 3956), Universit\'e Paris-Est, Senart-FB Institute of Technology, Lieusaint, France
              \email{aurelien.hazan@u-pec.fr}
              \and
              M. Olteanu \at
              SAMM (EA4543), Universit\'e Paris 1 Panth\'eon-Sorbonne, Paris, France\\
              MaIAGE, INRA, Universit\'e Paris-Saclay\\
              \email{Madalina.Olteanu@univ-paris1.fr}
              \and
              J. Randon-Furling \at
              SAMM (EA4543), Universit\'e Paris 1 Panth\'eon-Sorbonne, Paris, France\\
              \email{Julien.Randon-Furling@univ-paris1.fr} 
}

\date{Received: date / Accepted: date}

\maketitle

\begin{abstract}
We introduce a multidimensional, neural-network approach to reveal and measure urban segregation phenomena, based on the Self-Organizing Map algorithm (SOM). The multidimensionality of SOM allows one to apprehend a large number of variables simultaneously, defined on census or other types of statistical blocks, and to perform clustering along them. Levels of segregation are then measured through correlations between distances on the neural network and distances on the actual geographical map. Further, the stochasticity of SOM enables one to quantify levels of heterogeneity across census blocks.\\
We illustrate this new method on data available for the city of Paris.
\end{abstract}

\section{Introduction}

Various contexts can be characterized as reflecting one form or another of urban segregation. In most instances, the phenomenon is defined to be one where the distribution of a given variable locally (at the neighbourhood scale) differs substantially from its distribution at the scale of the whole city. A seminal study of the emergence of segregation was Schelling's checker-board simulation of a two-community dynamics~\cite{Schel1,Schel2} --~incidentally, one of the first multi-agent models. This model revealed how, even with relatively high levels of tolerance for mixity, a system of agents located at sites on a grid would eventually evolve towards patterns of segregation, with the two communities ``living'' in distinct parts of the grid. Schelling's model has given rise to a vast literature on segregation across a wide range of disciplines~\cite{Clark1,Crane1,Pollicott,Macy,Laurie,pancs,Clark2,Singh,Grauwin,henry,Banos,durrett2014}, and notably in statistical physics, where analogies with certain interacting particle systems have been put to fruitful use~\cite{Vinko,Stauffer,Dall,Gauv1,Castellano,Gauv2,RMK,HRF,cortez2015,sajen}. However, rare are the instances in which real-world data has been compared with theoretical results other than stylized facts~\cite{Bene1,Hatna1,Hatna2}. One reason for that is the notorious difficulty to elaborate, based on available data, segregation indices~\cite{reardon2004} that could correspond to some of the theoretical ones --~the evolution of which is described in agent-based models.

Starting from the data available for the city of Paris, we set out to provide a multidimensional picture of segregation phenomena, and to explore segregation indices that can be sensibly and robustly defined from this picture.

\begin{sloppypar}
Specifically, in this extended version of a communication given at WSOM+17~\cite{CORFH}, we report on the first steps towards the definition of a data-based segregation index \textit{via} Self-Organizing Maps~\cite{Kohonen1982,koh12}. SOM's intrinsic multidimensionality presents many benefits for the study of such a complex system as a city. This has only recently been noted~\cite{Arribas,Wei} and our work shall hopefully contribute to SOM becoming a standard tool in urban sociology and geography. All the more so as, most importantly and specifically, the topology obtained by SOM allows for useful comparisons with the actual geographical topology. Precisely, we show here (in section~4) how to measure segregation from correlations between SOM distances and geographical distances.
\end{sloppypar}

Beforehand, we describe in section~2 the pre-processing applied to the available public data, as well as the particular choice of variables we have opted for in this paper. In section~3, we apply SOM combined with hierarchical agglomerative clustering (HAC) on three sets of variables, and thus obtain three typologies of neighbourhoods with $4$ to $6$ well-identified archetypes for each set.

\section{Data and variables}
\begin{table}
\begin{center}
\begin{tabular}{|c|c|c|}
  \hline
  Set & Label & Variable \\
  \hline
 1 & \verb?Decile.1? & 1st decile of income dist.  \\
 \hline
 1 & \verb?Mediane? & Median of income dist.  \\
 \hline
 1 & \verb?Decile.9? & 9th decile of income dist.  \\
 \hline
 1 & \verb?Part_patrim? & Share of financial and patrimonial income  \\
 \hline
 1 & \verb?Part_min_soc? & Share of minimum social benefits  \\
 \hline
 2 & \verb?age_moy? & Average age \\
  \hline
 2 & \verb?std_age? & Standard deviation of age dist. \\
  \hline
 2 & \verb?moins_18_moy? & Average number of inhabitants under 18 y.o.  \\
  \hline
 2 & \verb?Diplom_moy? & Average level of education \\
 & & (1: pre-secondary; 5: postgrad.) \\
  \hline
 2 & \verb?std_diplom? & Standard deviation of education level \\
  \hline
 3 & \verb?taux_hlm? & Fraction of social housing \\
 \hline
 3 & \verb?1_D_EP_Sum? & Number of EP primary schools \\
 \hline
 3 & \verb?1_D_Public_Sum? & Number of state primary schools \\
 \hline
 3 & \verb?Col_EP_Sum? & Number of EP secondary schools \\
 \hline
 3 & \verb?Col_Public_Sum? & Number of state secondary schools \\
 \hline
 3 & \verb?1_D_Priv_Sum? & Number of non-state primary schools \\
 \hline
 3 & \verb?Col_Priv_Sum? & Number of non-state secondary schools \\
  \hline
 3 & \verb?Commerce_Sum? & Number of shops \\
  \hline
 3 & \verb?Services_Sum? & Post offices, local administration offices\\
  \hline
 3 & \verb?Sports_Sum? & Number of sports facilities \\
  \hline
 3 & \verb?Action_sociale? & Social services offices \\
  \hline
 3 & \verb?Medecins_Sum? & Number of medical doctors (GPs and specialists) \\
  \hline
 3 & \verb?Sante_Sum? & Number of hospitals, pharmacies\\
  \hline
 3 & \verb?Transport_Sum? & Number of train stations and travel agents\\
  \hline
 3 & \verb?Somme_lignes? & Number of metro and tram stops (within $800$m)\\
  \hline
\end{tabular}
\end{center}
\smallskip

\caption{Variables used for this exploratory study. EP stands for ``\'Education Prioritaire'', a state-sponsored scheme for schools in deprived neighbourhoods. (When counting shops, schools, services etc, the 10 nearest neighbouring blocks of a given IRIS are taken into account.)}
\label{tab:1}
\end{table}

Depending on its source and its type, available data comes in various formats. We used databases from INSEE (\textit{Institut National de la Statistique et des \'Etudes \'Economiques}, France's national agency for economical data), IGN (\textit{Institut G\'eographique National}, France's national agency for geographical data) and RATP (\textit{R\'egie Autonome des Transports Parisiens}, Paris public transport agency).

\subsection{Individual data}
The individual level is the household one. At this level, INSEE provides data on many characteristics such as the number and age of household members, education level of the household head, size and type of the house, etc.
We have aggregated data from this level to the census block level. This allows for instance to compute the fraction of social housing in a given census block.

\subsection{Census block data}
In INSEE data, census blocks are called IRIS (\textit{Ilots Regroup\'es pour l'Information Statistique}). Unfortunately, neither do they correspond to a fixed surface area nor to a fixed number of inhabitants, although they correspond on average to blocks of around $2,000$~inhabitants. Thus, the city of Paris, with a total population of over $2$~million people comprises just under $1,000$~IRIS. Some IRIS blocks are also purely geographical, with no or very few inhabitants. They appear nonetheless in INSEE data because of services and facilities they may offer. They also appear in the contour data provided by IGN to draw geographical maps and spatial representations.

At the census block level, INSEE provides data such as the number and types of shops, the number and types of public service offices, the number and type of health facilities. They also provide quantiles of the income distribution within each census block -- except those where the number of households is too small and combined with a high level of income. For this and other similar reasons, the number of IRIS for which data is available differs from one variable to the other. 

\subsection{Transport network}
The metropolitan authority for transportation provides geographical coordinates for all access points to underground, tramway and bus stations in the Paris area. Using this data (a version of it processed by the OpenStreetMap project~\cite{osm}), we have been able to compute for each census block the number of underground and tramway lines available within an~800 meter radius (from the centroid of the block). Note that a station with two lines counts twice as a station with just one line.

\subsection{Variables}
For this exploratory case study, we have retained three sets of variables for each census block:
\begin{itemize}
\item[Set 1] revenue and income: first and ninth deciles as well as median of the income distribution, fraction of revenue coming from assets and other patrimonial sources, fraction of revenue coming from minimal social benefits\footnote{These are social benefits paid to prevent people from falling into extreme poverty. They vary from $300$~euros to about $800$~euros per month.}. From the available data, these variables could be computed for 853 IRIS blocks in Paris.
\item[Set 2] population: age (average and standard deviation), number of people under 18 years old, education level (coded in 5 groups, average and standard deviation). From the available data, these variables could be computed for 943 IRIS blocks in Paris.
\item[Set 3] urban facilities and services: rate of social housing, access to public transport, number of shops, access to medical and health services, number of sports facilities, number of primary and secondary schools (including primary schools in special urban and education development projects -- called \textit{\'education prioritaire} (EP) in France). From the available data, these variables could be computed for 980 IRIS blocks in Paris.
\end{itemize}
A full list of the variables used in this study is given in Table~\ref{tab:1}.

\section{Self-Organizing Map Approach}

\begin{figure}[b]
\begin{center}
\subfigure[]{\includegraphics[scale=0.35]{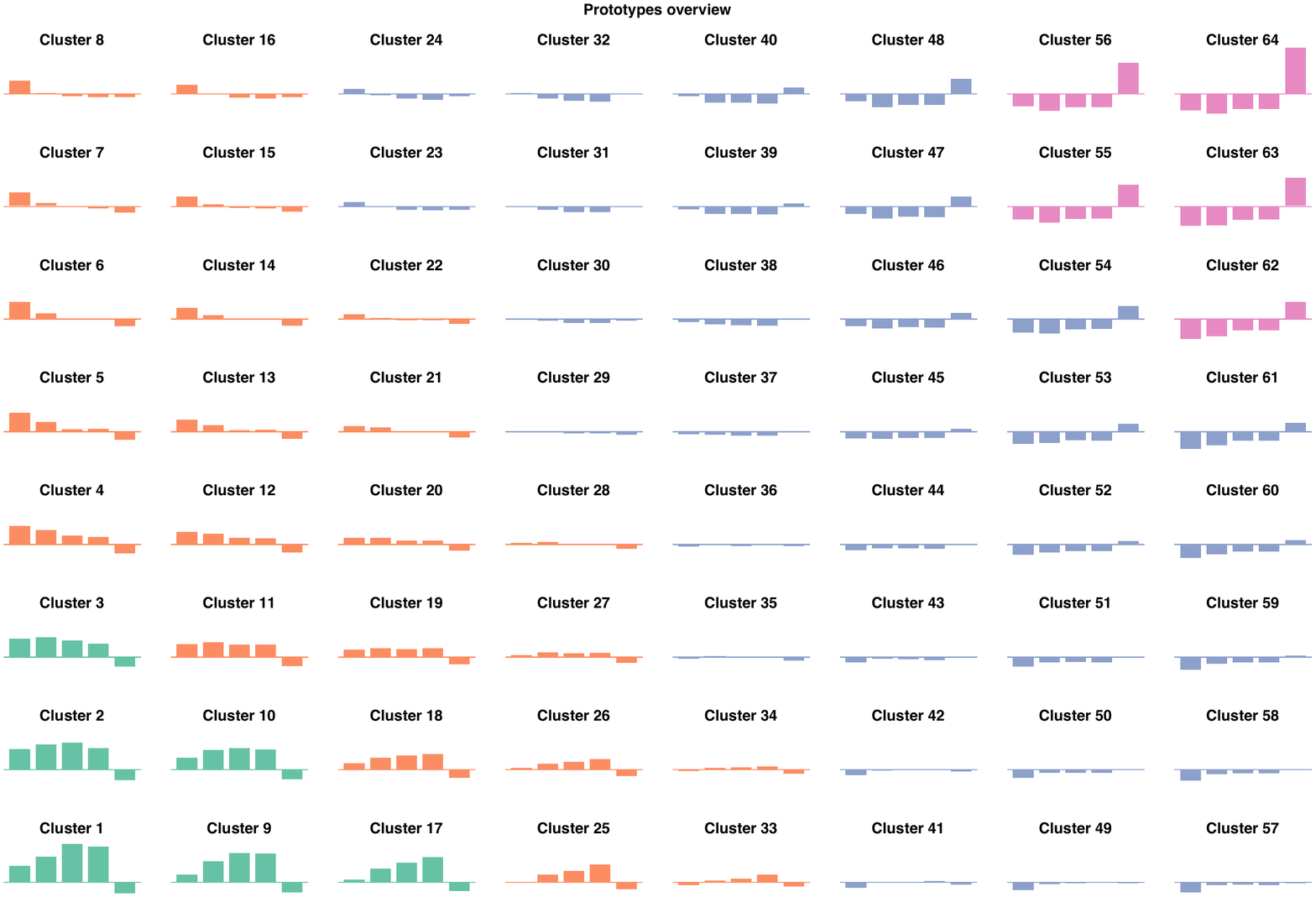}}
\end{center}
\caption{\label{fig:maps}Kohonen maps for the variables of Set~$1$~(subfigure~a), $2$~(b) and $3$~(c), with values of the variables for the prototypes in each class. Colours indicate groups of clusters obtained by hierarchical agglomerative clustering (HAC): 4 groups for Sets $1$ and $2$, 6 for Set $3$.}
\end{figure}
\begin{figure}
\ContinuedFloat
\begin{center}
\setcounter{subfigure}{1}
\subfigure[]{\includegraphics[scale=0.35]{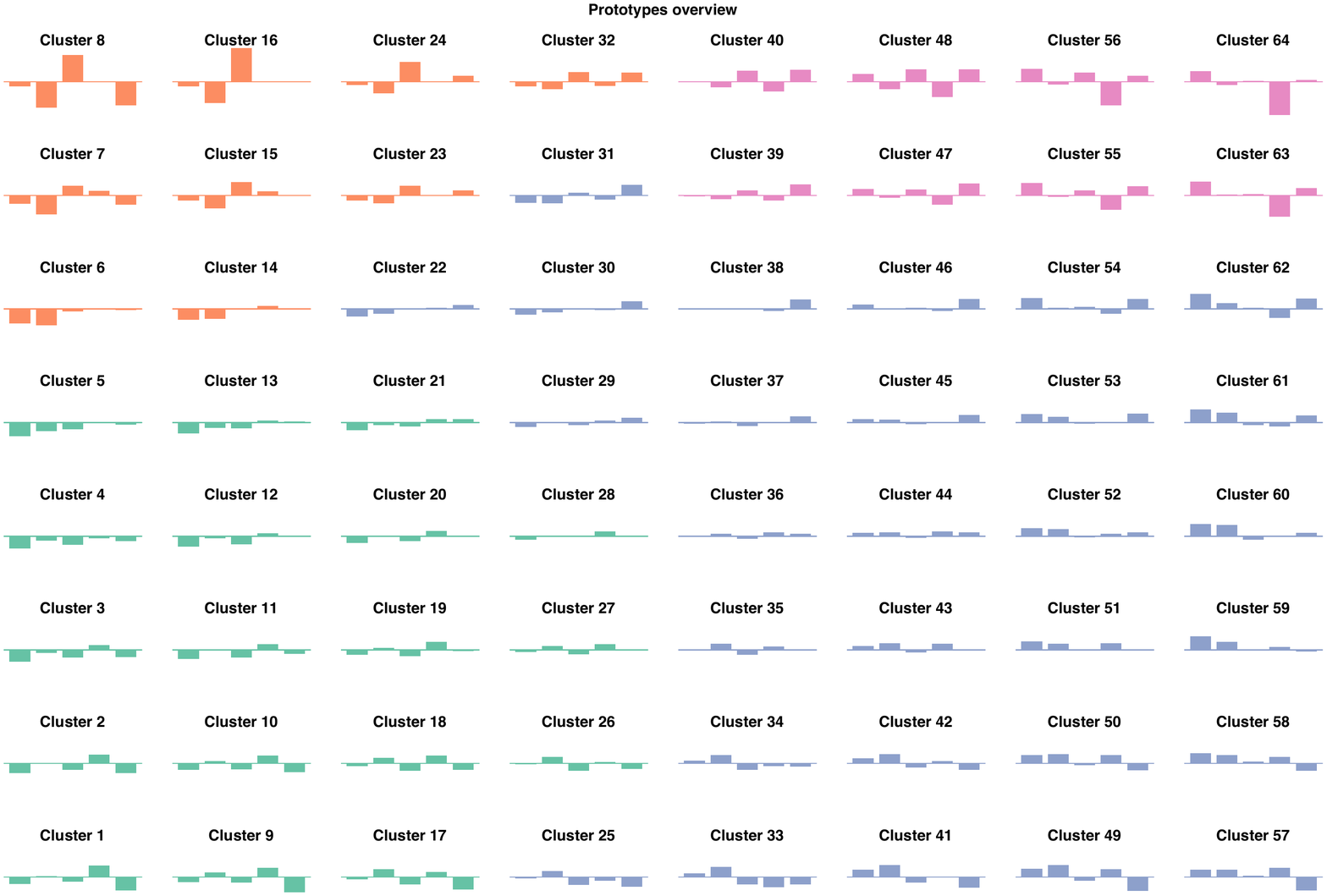}}
\subfigure[]{\includegraphics[scale=0.35]{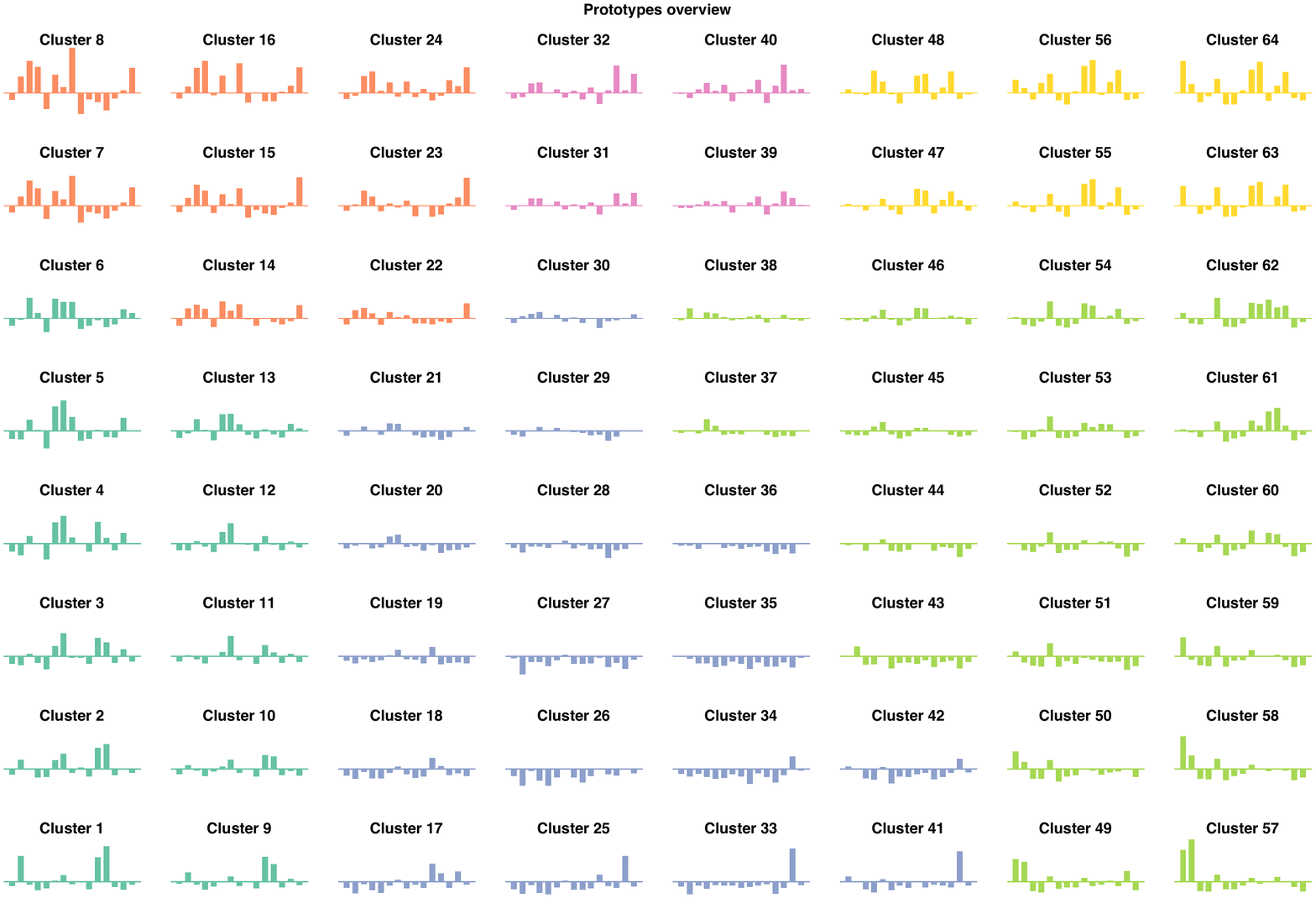}}
\end{center}
\caption{(continued) Kohonen maps for the variables of set~$1$~(subfigure~a), $2$~(b) and $3$~(c), with values of the variables for the prototypes in each class. Colours indicate groups of clusters obtained by hierarchical agglomerative clustering (HAC): 4 groups for Sets $1$ and $2$, 6 for Set $3$.}
\end{figure}

We use a multidimensional classification algorithm known as Self-Organizing Map and first introduced by T.~Kohonen~\cite{Kohonen1982,koh12}. All results were obtained using the R-package SOMbrero~\cite{SOMbrero}, which performs SOM combined with a hierarchical agglomerative clustering (HAC). Given that we are dealing with just under one thousand census blocks, we choose to train the algorithm to produce an 8x8~map for each of the three sets of variables, and then to classify individual IRIS blocks into four groups for sets $1$ and $2$ and six groups for the third set. The number of clusters is chosen from the dendrograms so as to provide a meaningful classification, with a balance between too little separation among groups and too few details in their description: four to six types of neighbourhoods compose a reasonably coarse-grained rendering of geographical and sociological details.

The online SOM method implemented in the SOMbrero package being a stochastic algorithm, we allowed for $100$~runs on each set of variables. We then extracted results from the runs exhibiting the best explained variance ratios. These are shown on Figures~\ref{fig:maps}, and we proceed to analyze them.

\begin{table}
\begin{center}
\begin{tabular}{|c|c|c|c|c|c|}
  \hline
  Group & 1st decile & Median income & 9th decile & Revenue & Revenue\\
  &  &  &  & from assets & from social benefits\\
  \hline
 1 & $13,405$ & $44,367$ & $129,538$ & $40$ & $0.2$ \\
 \hline
 2 & $12,504$ & $33,281$ & $73,929$ & $22$ & $0.5$\\
  \hline
 3 & $9,471$ & $23,963$ & $50,338$ & $13$ & $1.2$\\
  \hline
 4 & $7,390$ & $15,081$ & $30,840$ & $7$ & $3.6$\\
  \hline
 All & $10,672$ & $28,411$ & $65,309$ & $19$ & $??$ \\
  \hline
\end{tabular}
\end{center}
\smallskip

\caption{Per-group averages of some of the variables in Set~1 (in euros for quantiles of income distribution, in percent for the share of revenue drawn from financial and other assets).}
\label{tab:avSom1}
\end{table}

\subsection{SOM on Set 1}

\begin{figure}[b]
\begin{center}
\includegraphics[angle=-90,scale=0.35]{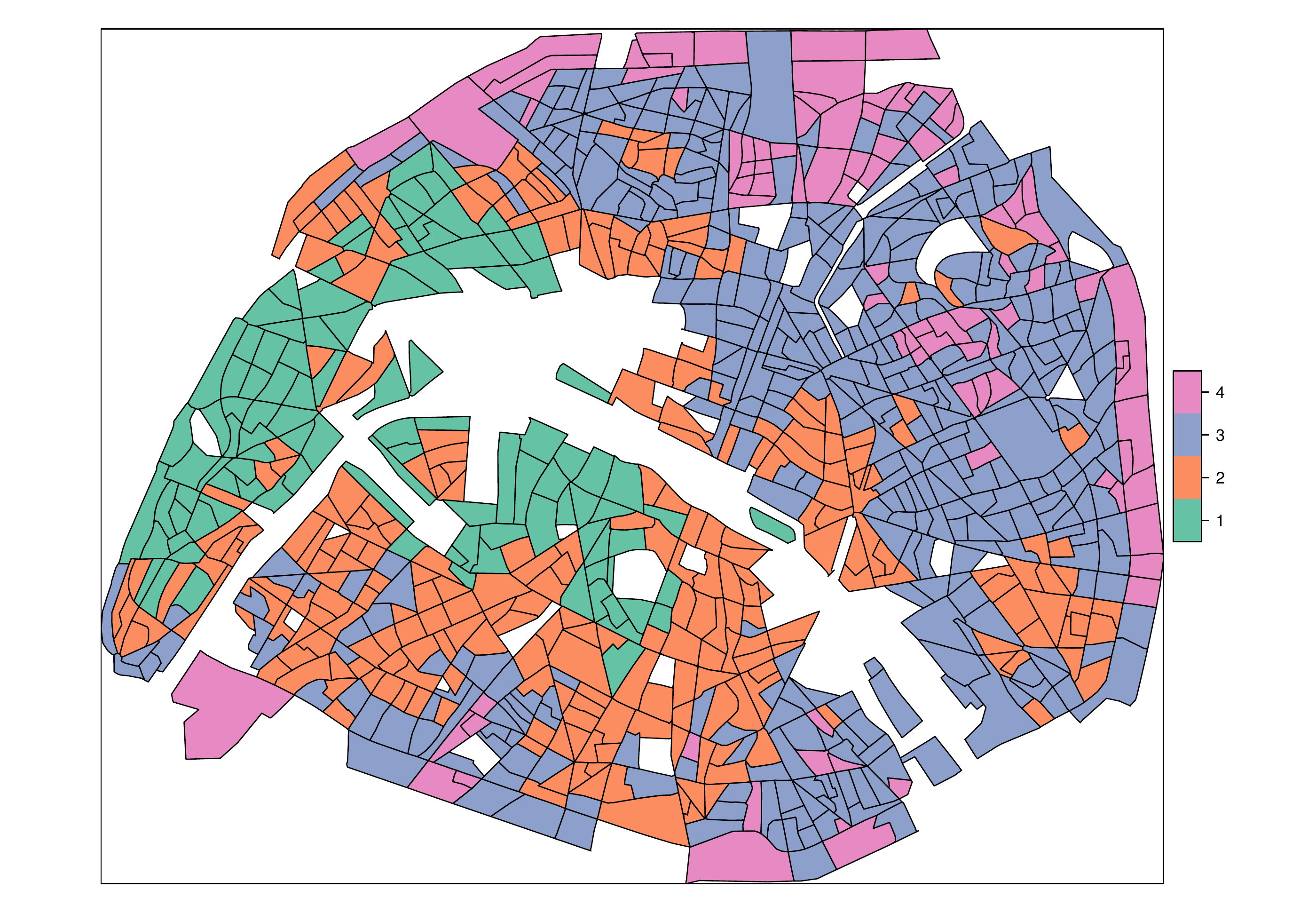}
\end{center}
\caption{\label{fig:ParisSom1}Spatial distribution of groups 1 to 4 for Set~$1$. (Colours correspond to groups as defined on Figure~\ref{fig:maps}. Areas in white correspond to parks, train stations and blocks for which data is not available.)}
\end{figure}

The first set of variables correspond to income and revenue variables. These are the most commonly used in socio-economic segregation studies (along with ethnic group variables -- let us recall here that ethnic statistics are not available in France). As can be seen on Figure~\ref{fig:maps}, SOM followed by HAC yields here well ordered groups, parallel to the diagonal, corresponding to four easily identifiable types of census blocks (see Table~\ref{tab:avSom1}). There are blocks (group~4) where the population is clearly poorer than the Parisian average, with a first decile of just above $7,000$~euros per year. At the other end of the spectrum, blocks where the population is not only richer than average (specifically the top $10$ percent are much wealthier, with a ninth decile above $129,000$ euros per year) but also with a very substantial part of revenues coming from financial and other patrimonial assets: $40$~percent on average (group~1). In between these two groups, the other two types of blocks correspond to upper (2) and lower (3) middle classes, with again a difference in the level of patrimonial income ($22$~percent vs $13$ percent).
Patrimonial income thus seems to work as an order variable (in the sense that it characterizes the cluster to which a block belongs). It is significantly correlated with spatial segregation: Figure~\ref{fig:ParisSom1} shows indeed a high level of spatial homogeneity for the groups derived from the first set of variables.

\subsection{SOM on Set 2}

\begin{table}
\begin{center}
\begin{tabular}{|c|c|c|c|c|}
  \hline
  Group & Age & Age & Children & Education\\
   & & std deviation & per household & \\
  \hline
 1 & 39.2 & 17.7 & 0.3 & 2.8 \\
 \hline
 2 & 39.3 & 14.4 & 0.6 & 2.7 \\
  \hline
 3 & 44.5 & 18.6 & 0.4 & 2.7 \\
  \hline
  4 & 46.4 & 16.7 & 0.5 & 2.1 \\
  \hline
  All & 42.6 & ?? & 0.4 & 2.6\\
  \hline
\end{tabular}
\end{center}
\smallskip

\caption{Per-group averages of some of the variables in set~2 (education level is from pre-secondary (1) to postgraduate (5) level).}
\label{tab:avSom2}
\end{table}

SOM followed by HAC run on the second set of variables yields a less structured clustering than in the case of the first set of variables (see Figure~\ref{fig:maps}). However, one can identify four groups and the following underlying trends: the bottom of the Kohonen map (groups $1$ and $3$) corresponds to blocks with fewer children and a higher education level than the top part (groups $2$ and $4$). Also group~$1$ has younger heads of households on average than group~$3$. Similarly in group~$2$ the population is comparatively younger than in group~$4$, and with a higher education level~(see Table~\ref{tab:avSom2}).

\begin{figure}[h]
\begin{center}
\includegraphics[angle=-90,scale=0.35]{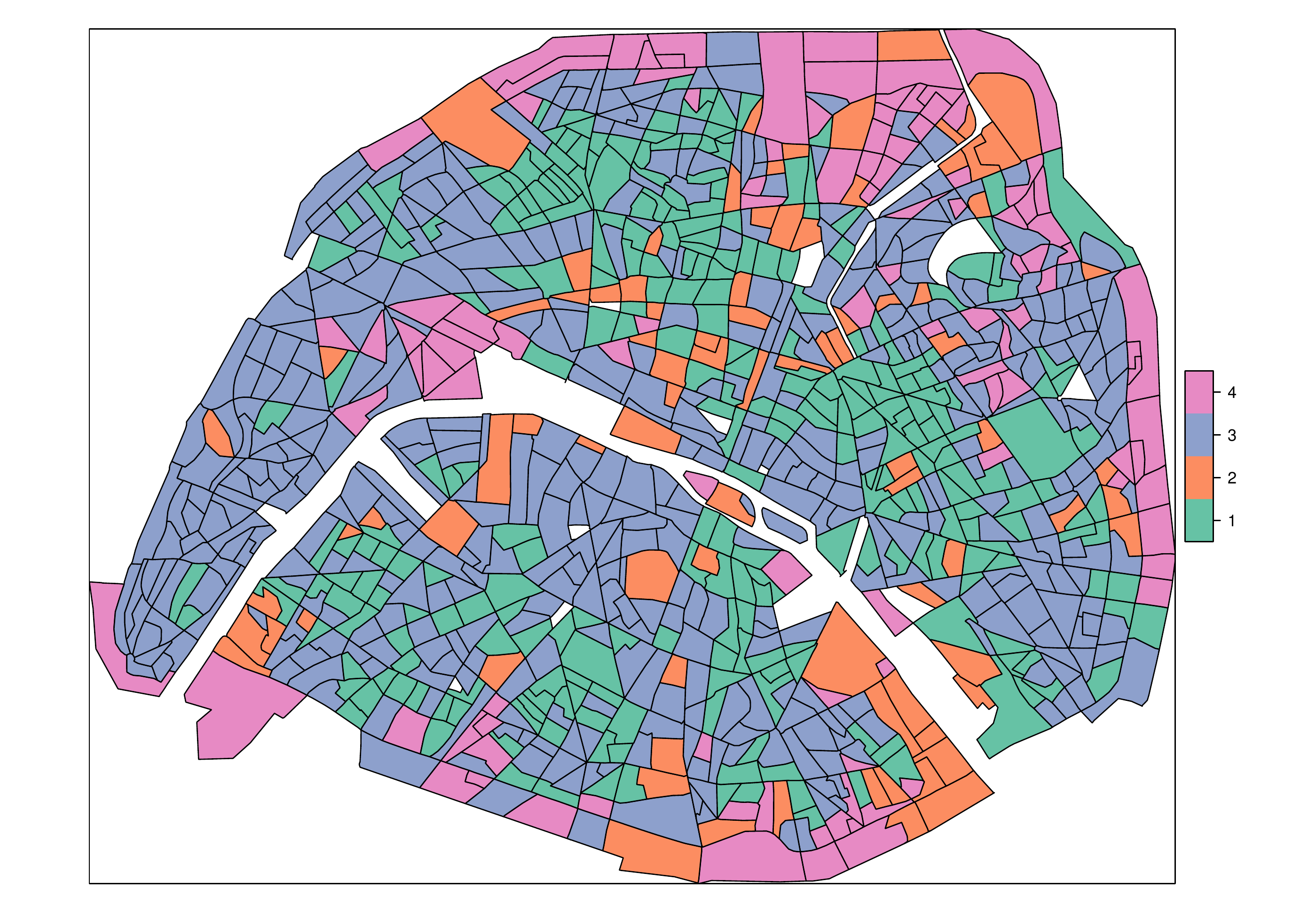}
\end{center}
\caption{\label{fig:ParisSom2}Spatial distribution of groups 1 to 4 for Set~$2$. (Colours correspond to groups as defined on Figure~\ref{fig:maps}.)}
\end{figure}

Spatially, groups have a wider distributions (Fig.~\ref{fig:ParisSom2}), but one still notes that certain areas are particularly representative of a given cluster, eg the northern-north-eastern part of the city for group~$4$.

\subsection{SOM on Set 3}
Processing the third set of variables with SOM and HAC allows one to distinguish six well-identified  types of census blocks, that pave the Kohonen map (Fig.~\ref{fig:maps} and Table~\ref{tab:avSom3}):
\begin{enumerate}
\item areas with more medical services, more private schools, fewer shops and less access to public transports;
\item areas with many shops, facilities and the highest access to public transports;
\item areas with a slight concentration of social housing ($9$~percent) and below average for all other variables;
\item areas with a high level of access to public transports, many shops and facilities and also a certain number of EP primary shools;
\item areas with a significant proportion of social housing ($34$~percent), very few EP primary schools, and the lowest access to public transport;
\item areas with the highest proportion of social housing ($42$~percent), the largest number of EP schools, and low access to public transport and other facilities.
\end{enumerate}

On this set of variables, a multidimensional approach such as ours sheds light on residential patterns, allowing to refine as one sees fit the level of description. For instance, in this case if one opts for only five groups, groups number~$5$ and $6$ will be merged, as may be seen from their proximity on the Kohonen map.

\begin{figure}
\begin{center}
\includegraphics[angle=-90,scale=0.35]{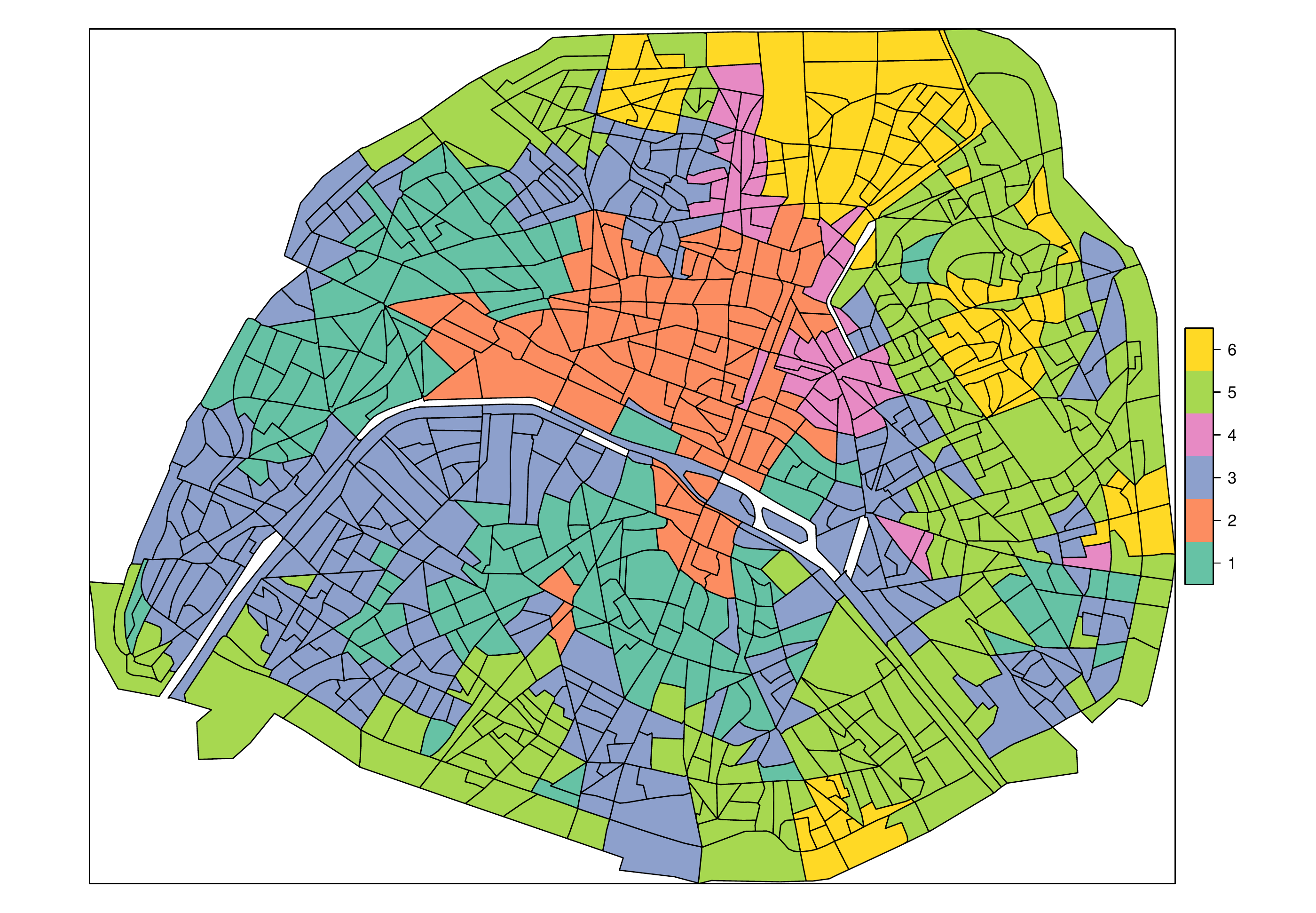}
\end{center}
\caption{\label{fig:ParisSom3}Spatial distribution of clusters 1 to 6 for Set~$3$. (Colours correspond to groups as defined on Figure~\ref{fig:maps}.)}
\end{figure}

If one looks at the spatial distribution of the groups (Fig.~\ref{fig:ParisSom3}), there are again some district areas (called \textit{Arrondissements} in Paris) that emerge as particularly representative of a given cluster: parts of the 5th, 6th and 16th Arrondissements for group~1, parts of the 1st, 2nd, 8th and 9th for group~2, parts of the 7th and 16th for group~3, the area around \textit{Place de la R\'epublique} for group~4, 13th, 19th and 20th for group~5, and north-eastern parts of the 18th for group~6.
\begin{table*}
\begin{center}
\begin{tabular}{|c|c|c|c|c|c|}
  \hline
  Group & Social housing & Medical doctors & EP schools & Shops & Public transport\\
  \hline
 1 & 5 & 298 & 0.1 & 362 & 7.2\\
 \hline
 2 & 5 & 181 & 1.4 & 626 & 16.6\\
  \hline
 3 & 9 & 172 & 0.5 & 236 & 7.2\\
  \hline
 4 & 10 & 129 & 5.4 & 406 & 12.4\\
  \hline
 5 & 34 & 129 & 2.8 & 177 & 5.9\\
  \hline
 6 & 42 & 100 & 10.5 & 182 & 6.3\\
  \hline
 All & 18 & 173 & 2.8 & 288 & 8\\
  \hline
\end{tabular}
\end{center}
\smallskip

\caption{Per-group averages of some of the variables in Set~3 (social housing rate is a percentage per block, other variables are raw numbers for each block and its ten nearest neighbours; access to public transport is the number of lines within $800$~meters of a block's centro\"id).}
\label{tab:avSom3}
\end{table*}

\section{Segregation indices}
Thanks to its multidimensional nature, the approach followed in the previous sections yields typologies of neighbourhoods according to multiple variables taken simultaneously into account. Now, for a given set of variables, are all types of neighbourhoods well mixed across the city, or are there any spatial patterns? In other terms, is there any form of spatial segregation along some of the variables considered here? Is there a set of variables for which segregation patterns are stronger than for the other two?\\
Looking at the maps on Figures~\ref{fig:ParisSom1}, \ref{fig:ParisSom2} and~\ref{fig:ParisSom3}, one sees spatial patterns --~but, how significant are they? We introduce in this section a method that allows to quantify them in a new manner, thanks to specific aspects of the SOM algorithm. 
\medskip

Indeed, a well-known difficulty arising in the study of segregation phenomena is that of defining indices to measure the actual level of segregation~\cite{batty1976,reardon2004,Feitosa}. Since most INSEE and other public data in Paris is only available at the IRIS level, let us think at this level. Then measuring segregation amounts to quantifying two things:
\begin{enumerate}
\item how different IRIS blocks are from one another (that is, individual census blocks are not all alike, eg some have a younger population than others, some have a richer population than others);
\item how much spatial concentration occurs for IRIS blocks of a given type (in terms of the SOM groups obtained in the previous section, this means how far one stands from a uniform distribution over the whole city for blocks belonging to each group).
\end{enumerate}

The first point may be addressed using the stochastic nature of the SOM algorithm. Indeed, if all blocks were (almost) identical for a given set of variables then the clustering obtained would not be very robust to a re-run of the algorithm with a different seed~\cite{debodt02}.~(See paragraph~\ref{sec:fickle} below.)

The second point may be treated from two perspectives. One can measure the geographical dispersion of blocks in a given cluster. Or, one can make use of one of SOM's specific properties. Indeed, proximity on the Kohonen map means proximity in the state space of the chosen set of variables. Thus comparing Kohonen distances and geographical distances gives a measure of segregation.
Based on this observation, we introduce ideas for a new segregation index. Beforehand, we compute some standard segregation indices.

\subsection{Some entropy and information theory indices}
A class of segregation indices is formed by the so-called entropy indices~\cite{theil1971,iceland2004}. Based on information theoretic entropy, they measure the difference between the entropy of the global distribution at the city scale and local distributions (we shall consider here each IRIS within its administrative neighbourhood --~with $91$ such neighbourhoods in Paris). A standard information theory index is the $H$~index defined in~\cite{reardon2002,reardon2004}, implemented in the R package \textit{seg}~\cite{Rseg}. Other standard segregation indices include the $R$-index (relative diversity) and the $D$-index (dissimilarity).

A summary of values obtained for these three standard segregation indices is given in Table~\ref{tab:seg}. All three indices indicates stronger segregation on the third set of variables, and weaker on the second one. The first set appears close to the third one in terms of segregation.

\begin{table}[h]
\begin{center}
\begin{tabular}{|c|c|c|c|}
  \hline
  Set of variables & Index 1 & Index 2 & Index 3\\
  \hline
 1 & 0.64 & 0.43 & 0.50 \\
 \hline
 2 & 0.44 & 0.22 & 0.28 \\
  \hline
 3 & 0.72 & 0.49 & 0.60 \\
  \hline

\end{tabular}
\end{center}
\smallskip

\caption{Values of various segregation indices for the clusters produced by SOM on our three sets of variables. Index~$1$ is the $D$-index. Index~$2$ is the $R$-index. Index~$3$ is the $H$-index. See~\cite{reardon2002,reardon2004} and the R package \textit{seg}~\cite{Rseg}}
\label{tab:seg}
\end{table}

\subsection{SOM-based segregation index}
As mentioned in the introduction of this section, a new integrated segregation index can be obtained by comparing Kohonen distances and geographical distances for all pairs of IRIS blocks.

\begin{figure}
\begin{center}
\centering
\subfigure[]{\includegraphics[scale=0.21]{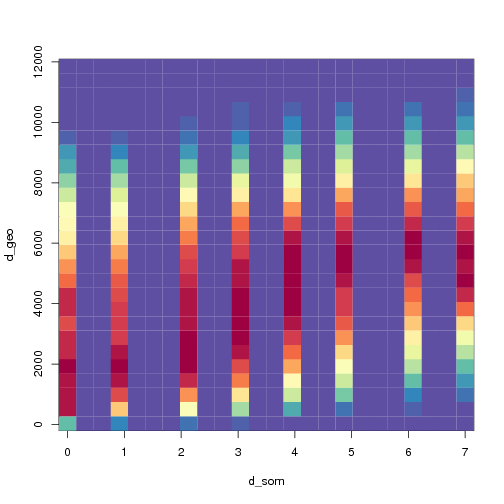}}\label{fig:entro2a}
\subfigure[]{\includegraphics[scale=0.21]{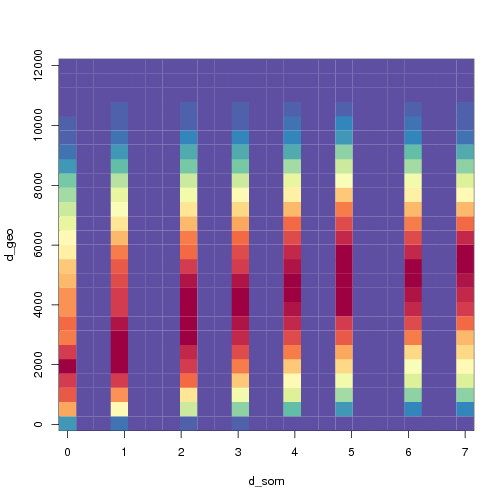}}\label{fig:entro2b}
\subfigure[]{\includegraphics[scale=0.21]{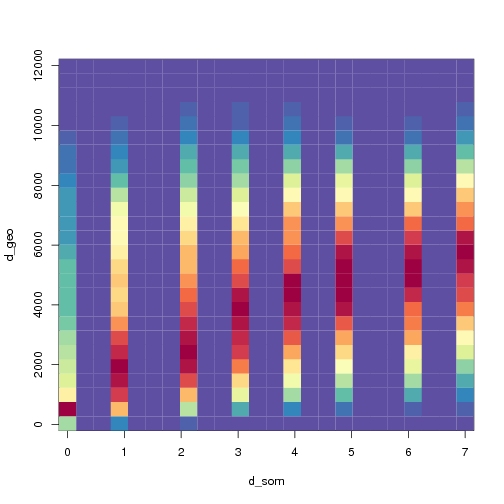}}\label{fig:entro2c}
\end{center}
\caption{\label{fig:entro2} Visual representation of a SOM-based segregation index for IRIS blocks considered with the variables of Sets~$1$ (a), $2$~(b) and $3$~(c). For each value of distance on the 8x8 Kohonen map, the density of pairwise geographical distances is shown (red means higher density). At a point with coordinates $(0,2000)$ one reads the probability density to find a pair of IRIS blocks that are in the same Kohonen class while their centroids are $2$~km apart in the actual city. If IRIS blocks from each Kohonen class were uniformly scattered across the city, geographical distance distributions would be the same for every Kohonen distance.}
\end{figure}

\begin{figure}
\begin{center}
\centering
\subfigure[]{\includegraphics[angle=-90,scale=0.13]{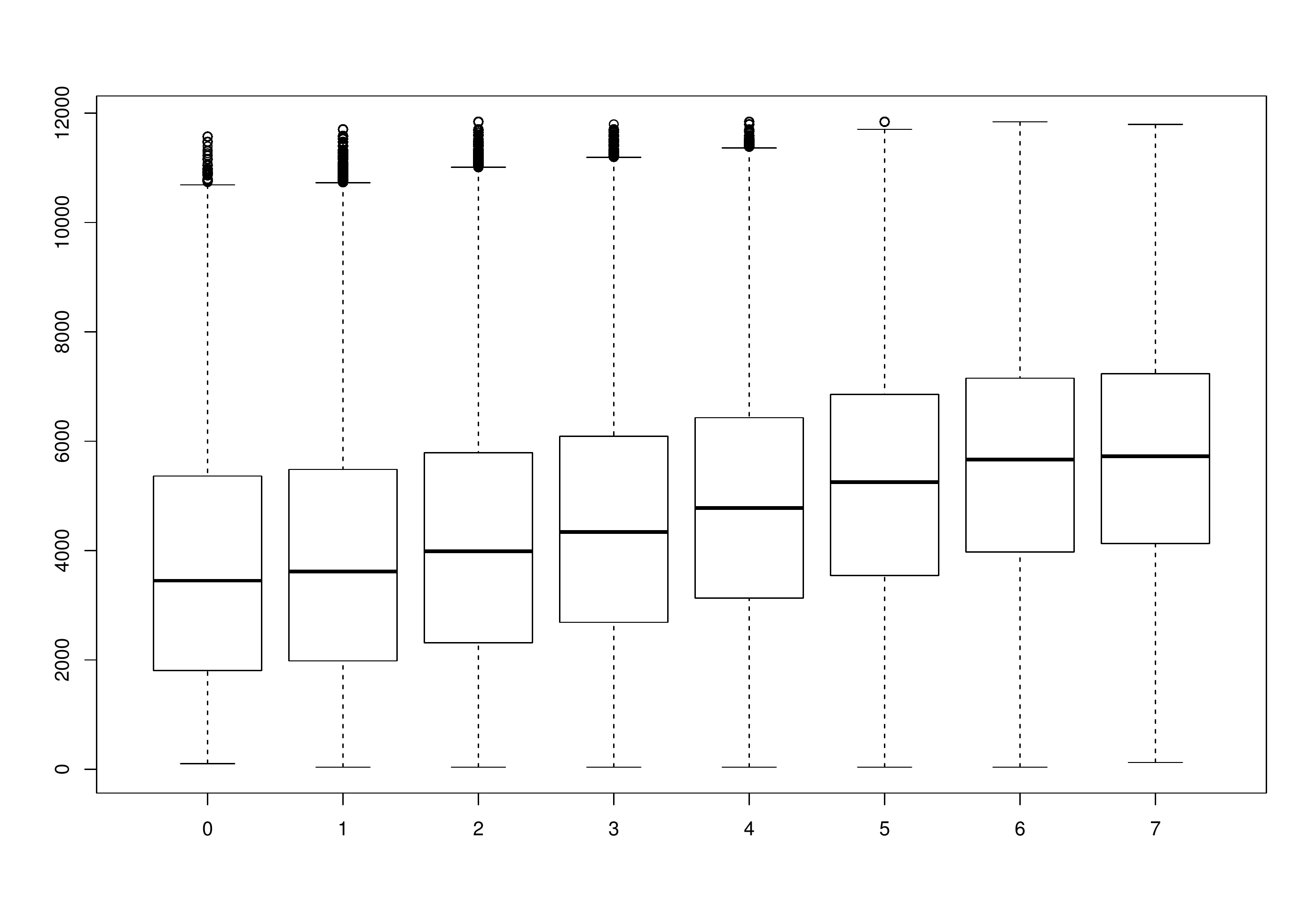}}
\subfigure[]{\includegraphics[angle=-90,scale=0.13]{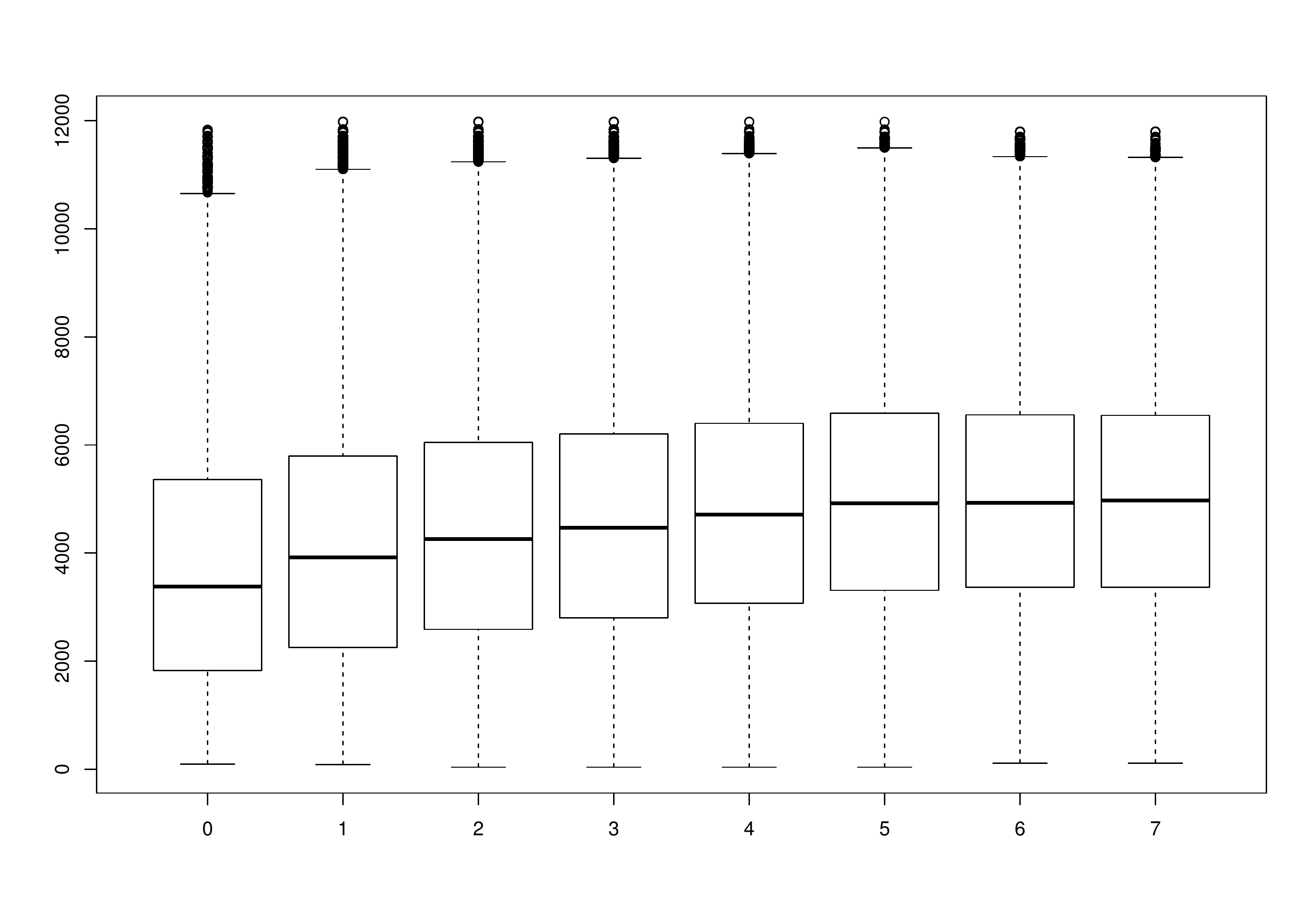}}
\subfigure[]{\includegraphics[angle=-90,scale=0.13]{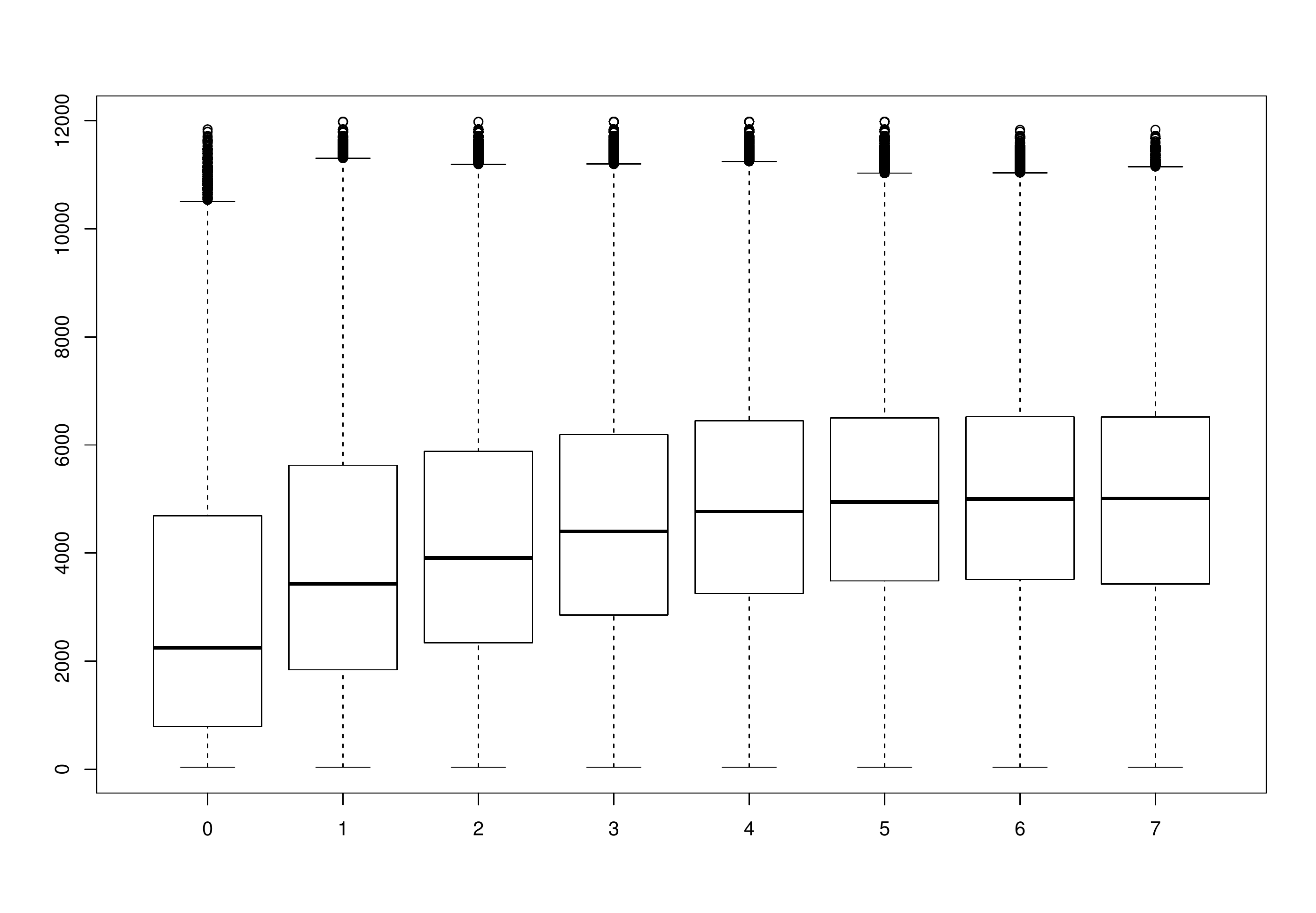}}
\end{center}
\caption{Box plot representations of the index presented on Figure~\ref{fig:entro2}.}
\label{fig:entro3}
\end{figure}

Indeed, proximity on the Kohonen map corresponds to proximity in the multidimensional space of the variables. The closer two blocks are on the Kohonen map, the more similar they are for the variables under consideration. Now, if the city were well mixed, one would not observe any particular spatial pattern: geographical distances would be independent of Kohonen distances. Thus any correlation between geographical and Kohonen distances signals spatial patterns, i.e. the presence of segregation, the level of which is well quantified by the actual value of the correlation.

One convenient way of visualizing this correlation is to represent the density of pairwise geographical distances for each value of the Kohonen distance, as on Figures~\ref{fig:entro2} and~\ref{fig:entro3}. Were the city well mixed, geographical distance distributions would be the same for every Kohonen distance\footnote{Note that such a simple, direct measure of the correlation between geographical distance and Kohonen distance is well suited to intricate patterns of segregation, as observed in real cities. However, if one considers artificial patterns with much regularity, this correlation measure works well on checkerboard patterns (provided the mesh is not too small), but obviously not as well on concentric patterns. More work is needed to circumvene this difficulty.}. This is not the case here, revealing in particular greater levels of segregation for the variables of Set~1, contrary to what appeared with standard segregation indices (see Table~\ref{tab:seg}).

Indeed, looking at numerical values (see Table~\ref{tab:SOMseg}), one observes that the correlation between Kohonen and geographical distances is twice as large for Set~1 as for Set~2, and about one-and-a-half as large for Set~1 as for Set~3.\\
Numerical values are to be compared with similar correlation coefficients computed in imaginary, extreme cases. For instance, if one considers four groups fully separated (with four ``pure'' neighbourhoods on the four quadrants of a square city), the correlation may be computed exactly and is about $0.61$. Given this value for the most extreme (unrealistic) case, one may say that the city of Paris exhibits significant levels of spatial segregation on all three sets of variables.

\begin{table}[h]
\begin{center}
\begin{tabular}{|c|c|}
  \hline
  Set of variables & SOM-based segregation index\\
  \hline
 1 & 0.26 \\
 \hline
 2 & 0.13 \\
  \hline
 3 & 0.18 \\
  \hline

\end{tabular}
\end{center}
\smallskip
\caption{Values of the SOM-based segregation index on the three sets of variables. This is defined as the correlation between Kohonen distances and geographical distances.}
\label{tab:SOMseg}
\end{table}

\subsection{Quantifying the heterogeneity of IRIS blocks\label{sec:fickle}}
\begin{figure}[t]
\begin{center}
\centering
\subfigure[]{\includegraphics[angle=-90,scale=0.12]{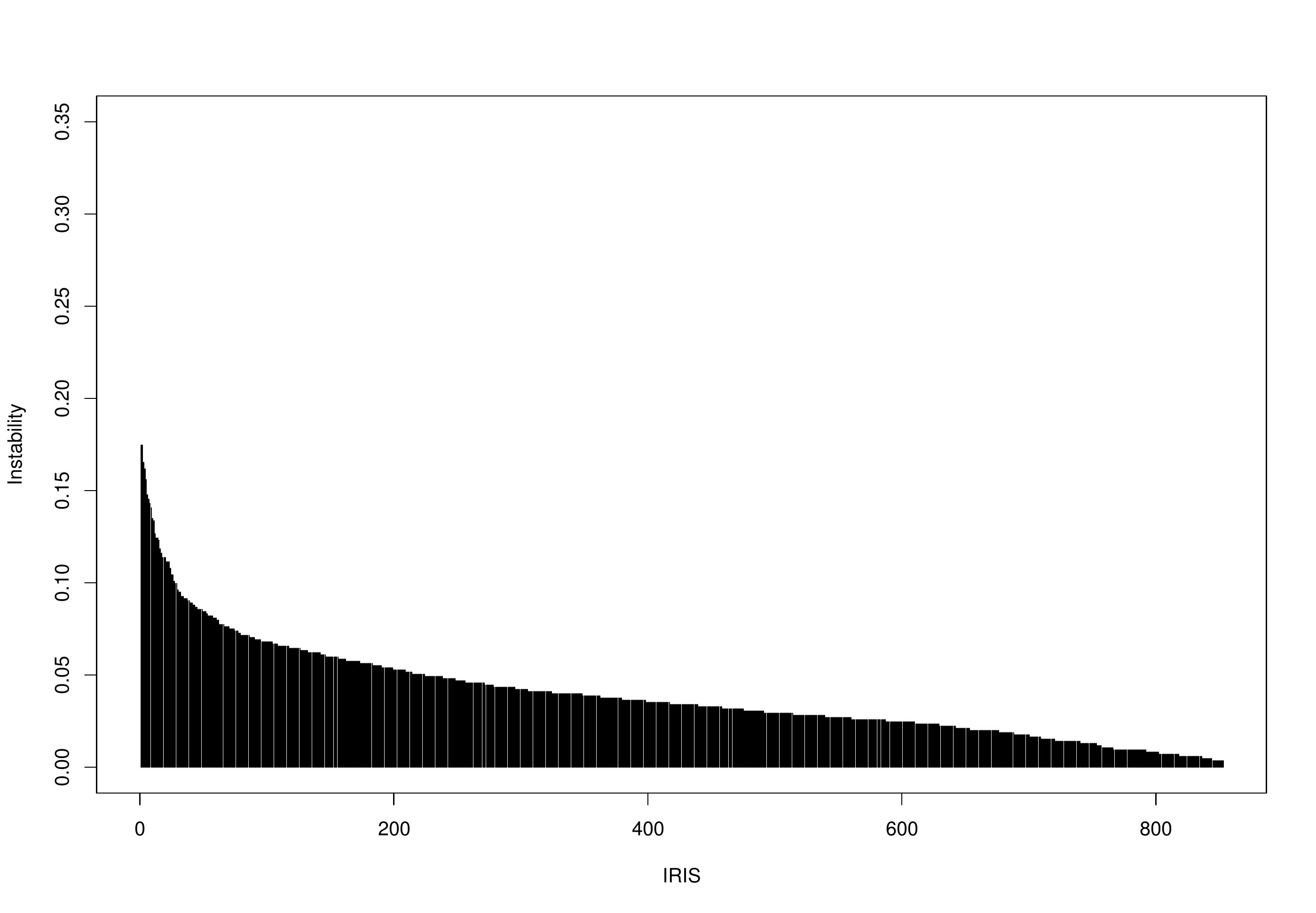}}
\subfigure[]{\includegraphics[angle=-90,scale=0.12]{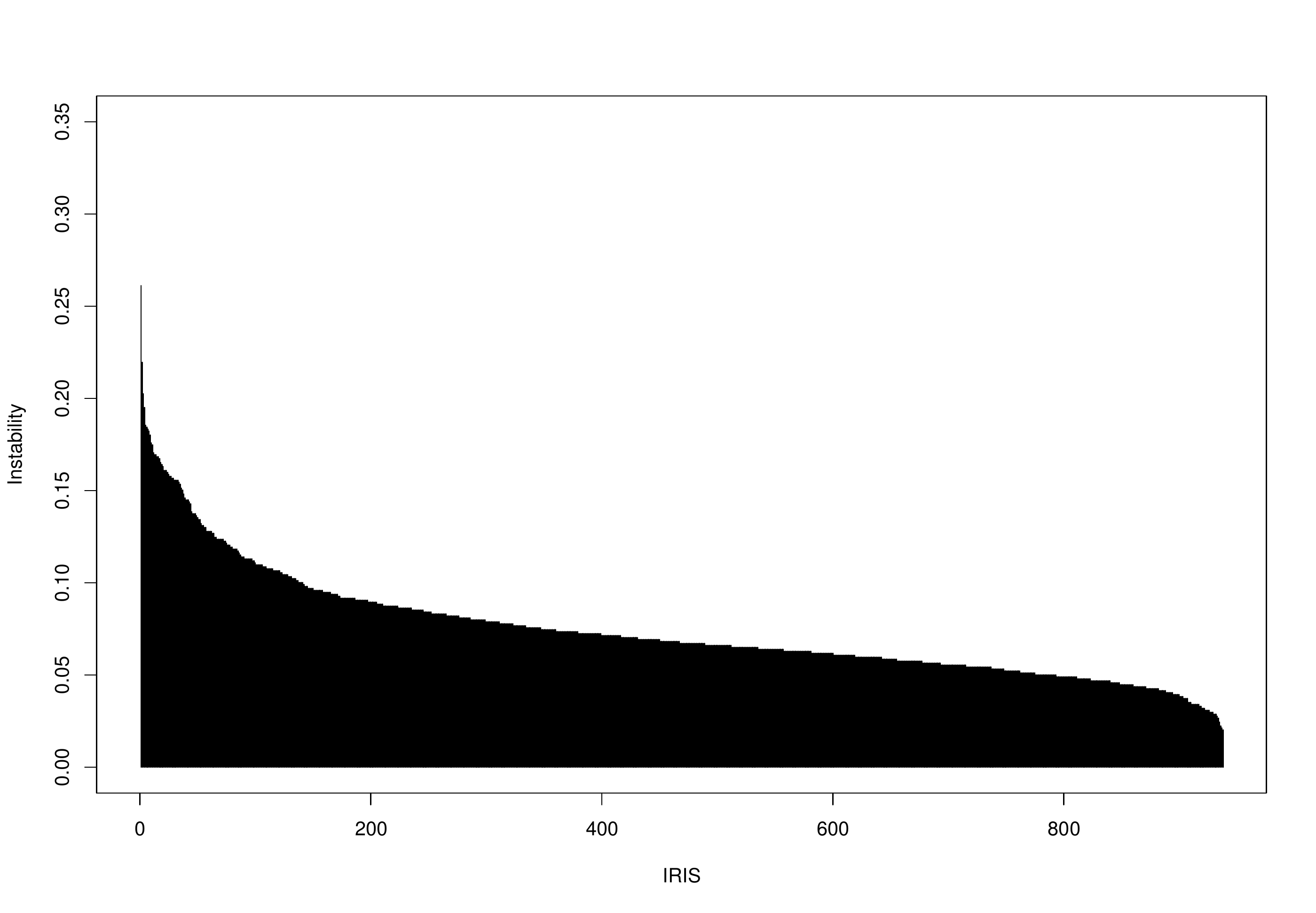}}
\subfigure[]{\includegraphics[angle=-90,scale=0.12]{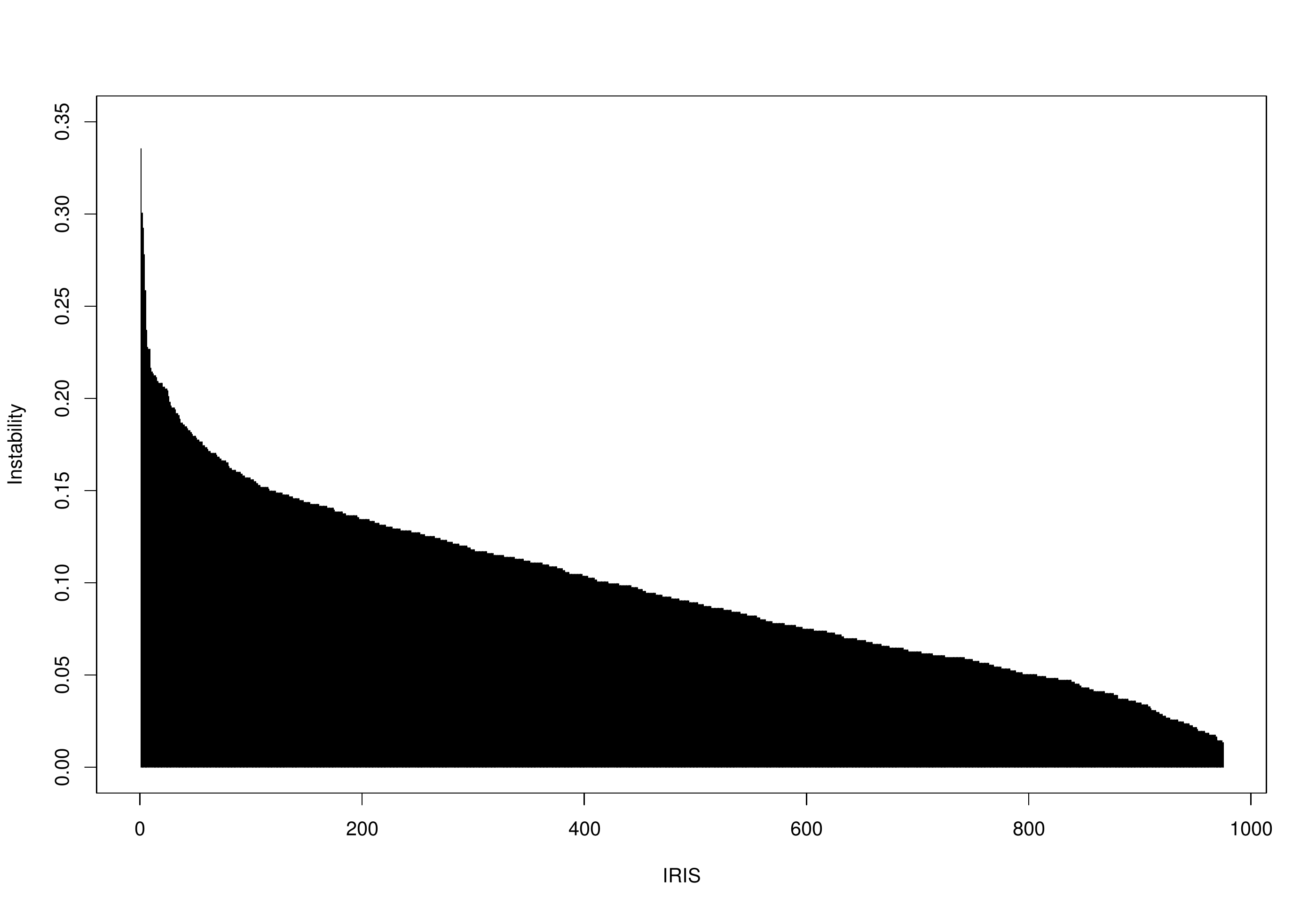}}
\end{center}
\caption{\label{fig:FiPa}Volatility part of the SOM-based segregation index on each of the three sets of variables defined in Table~\ref{tab:1}. Plots show level of fickleness for each IRIS block, in decreasing order of magnitude, for Sets~1 (left), 2 (centre) and 3 (right).}
\end{figure}

Any comparison between geographical distances and the distances obtained through a clustering algorithm obviously relies on the robustness of the classification. One should therefore control for the volatility level of the clustering: for a given pair of IRIS blocks, how sure can one be that they lie at a given Kohonen distance?
The stochastic nature of the SOM algorithm allows one to address this question. Indeed, volatility can be measured by looking at the fraction of pairs of areal units that are stable --~i.e. that always belong together to the same cluster (or the same vicinity) through a large number of SOM runs launched with different, random seeds~\cite{debodt02,bour15,bour15b}. One may then count, for each unit, the number of unstable pairs to which it belongs, thus defining its level of ``fickleness''.\\
As an illustration, we show on Figure~\ref{fig:FiPa} levels of fickleness for the three sets of variables considered in the previous sections of this paper. SOM classification appears robust across all three sets of variables, with a maximum of $10\%$ of fickle pairs observed on Set~3. Also, the classification is more robust on Set~1 than on Sets~2 or 3, indicating a greater level of differentiation among blocks along the variables of Set~1.

\begin{table}[h]
\begin{center}
\begin{tabular}{|c|c|}
  \hline
  Set of variables & Percentage of fickle pairs\\
  \hline
 1 & 4.0 \\
 \hline
 2 & 7.5 \\
  \hline
 3 & 9.6 \\
  \hline

\end{tabular}
\end{center}
\smallskip
\caption{Percentage of unstable pairs computed on $100$~runs of the SOM algorithm for each set of variables. An unstable pair is defined as a pair of IRIS blocks that do not tend to be always in the same vicinity or always \textit{not} in the same vicinity on the Kohonen map through the $100$~runs.}
\label{tab:Fick}
\end{table}

The larger the number of fickle units, the less heterogeneity there is among them and therefore the less definite will any classification be. In the extreme case in which all areal units present the same characteristics, any choice of classes will be arbitrary and this will be reflected in the fact that every pair of units will be unstable and every unit will be fickle. Note that this is not only a measure of the robustness of the classification: it actually measures the level of heterogeneity between areal units, independently of their spatial distribution. In this respect, levels of fickleness form a second part of a SOM-based segregation index (the volatility part). They complement the correlation defined in the previous paragraph, which measures the spatial aspect of segregation. \\

\section{Conclusion and perspectives}

The use of self-organizing maps to explore socio-economical and geographical data presents many promising features. The method is intrinsically multidimensional, and the clustering obtained is very much suited to interdisciplinary work: for instance, typologies of urban areas infered from sociological surveys and other forms of studies can be compared to clusters obtained from SOM, and their robustness can be tested by performing a large number of runs with different seeds.\\
Further, this new method allows for the definition of a general, robust segregation measure, both on the social and spatial levels --~the former is measured by the percentage of fickle pairs, and the latter by the correlation between Kohonen and geographical distances. We are working onward to develop a full methodological framework, with a formal definition and a thorough study of the new segregation index introduced here.

\section*{}
\noindent\textbf{Conflict of Interest}\\
The authors declare that they have no conflict of interest.
\bibliographystyle{spmpsci}

\end{document}